\documentclass[twocolumn,showpacs,prl]{revtex4}
\usepackage{amssymb}

\usepackage{amsmath}
\usepackage{epsf}

\begin{document}

\title{Cooper channel and the singularities in the thermodynamics of a Fermi liquid}
\author{Andrey V. Chubukov$^1$ and Dmitrii L. Maslov$^2$}

\begin{abstract}
We analyze how the logarithmic renormalizations in the Cooper channel affect
the non-analytic temperature dependence of the specific heat coefficient $%
\gamma (T) - \gamma (0)=A(T)T$ in a 2D Fermi liquid. We show that
$A(T)$ is expressed exactly in terms of the fully renormalized
backscattering amplitude which includes the renormalization in the
Cooper channel. In contrast to the 1D case, both charge and spin
components of the backscattering amplitudes are subject to this
renormalization. We show that the logarithmic renormalization of
the charge amplitude vanishes for a flat Fermi surface, when the
system becomes effectively one-dimensional.
\end{abstract}

\affiliation{$^1$Department of Physics, University of
Wisconsin-Madison, 1150 Univ. Ave., Madison, WI 53706-1390}

\vspace{0.5cm}%
\affiliation{$^2$Department of Physics, University of Florida, P.
O. Box 118440, Gainesville, FL 32611-8440} \maketitle Non-analytic
behavior of the thermodynamic parameters of a Fermi liquid has a
been of interest since the early days of Fermi-liquid theory, when
it was found that for $D=3$ the specific heat coefficient $\gamma
(T)=C(T)/T$ has a non-analytic temperature dependence $\delta
\gamma (T)=\gamma (T)-\gamma (0)\propto T^{2}\ln T$~\cite{FL}. The
issue of non-analyticities has been revived in recent years,
following the work by Belitz, Kirkpatrick, and Vojta
~\cite{belitz}, who found that the same type of logarithmic
behavior holds for the spin susceptibility at a finite momentum
$q$ in three dimensions (3D) $\delta \chi _{s}(q)\propto q^{2}\ln
q$.

Non-analyticities are stronger in two than in three dimensions: $\delta \gamma (T,H)$ and $%
\delta \chi _{s}(T,q,H)$ are linear functions of their respective
variables \cite{all_1,all_2,all_3,cm,all_4,efetov,finn,we,cmgg,cmm,chm,comm_a,ae}. In Refs.
\cite{cmgg,cmm}, it was argued that the non-analytic part $\delta
\gamma \left( T\right) $ comes from one-dimensional scattering
processes embedded in a 2D phase space~\cite{comm_a} and that
$\delta \gamma (T)$ is expressed via the charge and spin
components of the exact scattering amplitude with zero total
momentum $f(\theta =\pi )$ (``backscattering amplitude'')
\begin{equation}
\delta \gamma (T)=-\frac{3\zeta (3)}{2\pi \left( v_{F}^{\ast }\right) ^{2}}~%
\left[ f_{c}^{2}(\pi )+3f_{s}^{2}(\pi )\right] T.  \label{jul3_4}
\end{equation}
Validity of Eq.(\ref{jul3_4}) was verified perturbatively in Ref.\cite{cmgg}
by calculating $\delta \gamma (T)$ and $f_{c,s}(\pi )$ independently to
third and second order in the interaction, respectively, and checking that
Eq.(\ref{jul3_4}) holds.  That analysis, however, was incomplete -- it included the
renormalizations in the particle-hole channel but neglected the
renormalizations in the Cooper channel. Refs. \cite{cmgg,cmm}
conjectured--without proof-- that Eq.(\ref{jul3_4}) still holds if the
Cooper renormalizations are included into $f_{c,s}(\pi )$.

The Cooper renormalization of the backscattering amplitudes is an
essential ingredient of the theory, particularly in the limit
$T\rightarrow 0$. The argument is that the backscattering
amplitude \[f_{\alpha \beta ;\gamma \delta }(\pi )=f_{c}(\pi
)\delta _{\alpha \beta }\delta _{\gamma \delta
}+f_{s}(\pi )\mathbf{\vec{\sigma}}_{\alpha \beta }\cdot \mathbf{\vec{\sigma}}%
_{\gamma \delta }\] is expressed in terms of two full vertices
with zero
total momentum and either zero or $2k_{F}$ momentum transfer $\Gamma (%
\mathbf{k,-k;k,-k})\equiv \Gamma (\mathbf{k,k})$ and $\Gamma (\mathbf{%
k,-k;-k,k})\equiv \Gamma (\mathbf{k,-k})$:
\begin{equation}
f_{c}(\pi )=\frac{m}{\pi }\left[ \Gamma (\mathbf{k,k})-\frac{1}{2}\Gamma (%
\mathbf{k,-k})\right] ,f_{s}(\pi )=-\frac{m}{2\pi }\Gamma (\mathbf{k,-k})
\label{c_1}
\end{equation}
Equivalently, each of these two vertices can be expressed via the
fully renormalized
Cooper vertex with either zero or $2k_{F}$ momentum transfer: $\Gamma (%
\mathbf{k,k})=\Gamma ^{C}(0),~\Gamma (\mathbf{k,-k})=\Gamma
^{C}(2k_{F})$. Therefore, these vertices can be expressed via the
partial components $J_{n}^{C}$ of the irreducible interaction in
the Cooper channel as~\cite {landau}
\begin{equation}
\Gamma ^{C}(0),\Gamma ^{C}(2k_{F})=\sum_{n}(\pm 1)^{n}\frac{J_{n}^{C}}{1+%
\frac{m}{2\pi }J_{n}^{C}\ln {\frac{E_{F}}{T}}}.  \label{c_2}
\end{equation}
At low temperatures, all terms in the sum scale as $(\ln
E_{F}/T)^{-1}$, so that $\Gamma (k,k)$ and $\Gamma (k,-k)$ should,
in general, be reduced by the same logarithmic factor. Whether
this logarithmic renormalization affects 
 the backscattering amplitudes and $\delta \gamma (T)$ 
 is a more subtle issue
which is to be addressed by a direct calculation.

 The 1D case serves as a good example here. Both
$\Gamma (k,k)$ and $\Gamma (k,-k)$ are  logarithmically
renormalized in 1D, yet, these renormalizations 
only affect the spin channel, but 
cancel out in the charge
channel\cite{bychkov,review,emery,schulz_95}. As a result, the specific
heat remains linear in $T$ at the lowest $T$. This agrees 
with 1D bosonization according to which
 charge excitations  are described by a free Gaussian
theory, whereas the spin channel for the case a repulsive interaction
contains a marginally
irrelevant 
 perturbation that causes a logarithmic flow of the spin
amplitude. The issue that we address in this paper is whether this
situation occurs only in 1D or in higher dimensions as well.

The interplay between the logarithmic renormalization of the
interaction and the behavior of specific heat coefficient in 2D
has recently been considered by Aleiner and Efetov (AE)~\cite{ae}
(for a subsequent analysis see ~\cite {efetov,finn}). AE invented
an elegant supersymmetric method to treat the problem in arbitrary D by
integrating out fermions and expressing the low-energy action
solely in terms of the low-energy collective bosonic modes.
 They found that the spin contribution to $\delta\gamma(T)$ is
affected by the Cooper renormalization and behaves as $T(\ln |\ln
T|/\ln T)^{2}$ in the limit of $T\rightarrow 0$. They treated the 
 charge component in the eikonal approximation which neglects 
the curvature of the Fermi surface, and found that, 
in this approximation, the charge component 
 remains unrenormalized.

In this communication, we re-consider this issue. We evaluate $\delta \gamma
(T)$ explicitly to third order in the interaction $U(q),$ including Cooper
renormalizations, and also evaluate the spin and charge components of the
backscattering amplitude to second order in $U(q)$. We find that, in
contrast to the 1D case, \emph{both} $f_{s}(\pi )$ and $f_{c}(\pi )$ undergo
logarithmic renormalization in 2D. We also find that Eq. (\ref{jul3_4})
still holds when $U(q)$ is re-expressed in terms of $f_{s}(\pi )$ and $%
f_{c}(\pi )$. This result agrees with the conjecture made in Refs.
\cite {cmgg,cmm}. For a short-range interaction, we find that the
spin contribution to $\delta \gamma (T)$ scales as $(1/\ln
T)^{2}$, while the charge contribution scales as $(\ln |\ln T|/\ln
T)^{2}$. As a result, if the system remains in the normal state
down to \textit{very} low $T$, $\delta \gamma (T)$ would be
dominated by charge fluctuations and behaved as $\delta \gamma
(T)\propto T(\ln |\ln T|/\ln T)^{2}$. The normal state behavior in
2D, however, exists only down to the temperature $T_{p}$ of the
Kohn-Luttinger instability towards $p-$wave
pairing~\cite{kl,chub_kl}. We analyzed $\delta
\gamma (T)=A(T)T$ near $T_{p}$ and found that $A(T)$ diverges as $%
(T-T_{p})^{-2}$.

As a simple check, we verified that the same procedure that we
used for calculating the scattering amplitudes in 2D reproduces
the known result in 1D, namely, the cancellation of the
logarithmic renormalizations in the charge
channel\cite{bychkov,review,emery,schulz_95}. This comparison
helps to see where exactly the cases of $D>1$ and $D=1$ differ:
for $D>1,$ only the Cooper channel is logarithmic; hence there is
no cancellation between the Cooper and particle-hole channels, and
$f_{c}\left( \pi \right) $ is renormalized along with $f_{s}\left(
\pi \right) .$ For $D=1$, both particle-hole and particle-particle
renormalizations are logarithmic, and the two cancel each other in
the charge channel. To further emphasize this point, we also
considered the 2D case with a non-circular Fermi surface and
demonstrated that when the Fermi surface becomes flat, the
particle-hole renormalization becomes logarithmic and the Cooper
and particle-hole renormalizations again cancel each other in the
charge channel. 

We believe that it is the effect of the curvature that is responsible
for the difference between our result and that of AE.

We begin with the 2D case and a circular Fermi surface.
To second order in $U(q)$, the diagrams for the thermodynamic
potential $\Xi $ contain two fermionic bubbles $\Pi (q,\Omega )$.
To this order, ~the non-analyticity comes from the square of the
\textit{dynamic} part of $\Pi (q,\Omega )$ which contains the term
$\Omega ^{2}/q^{2}$ at $v_{F}q\gg |\Omega |$ \cite{cmgg,cmm,finn}.
Because the momentum integral is logarithmic in 2D, the
second-order thermodynamic potential
\begin{equation}
\Xi \sim U^{2}T\sum_{\Omega }\int d^{2}q\Pi ^{2}\sim U^{2}T\sum_{\Omega
}\Omega ^{2}\ln |\Omega |\sim U^{2}T^{3}  \label{dm1}
\end{equation}
contains a universal $T^{3}$ term, which gives rise to a $O(T)$ term in $%
\delta \gamma $. It was shown in~Refs. \cite{cmgg,cmm,finn} that the momenta
carried by fermions in the two bubbles, $\mathbf{k}_{1},\mathbf{k}_{2},$ $%
\mathbf{k}_{3},$ and $\mathbf{k}_{4},$ are correlated in such a way that $%
\mathbf{k}_{1}\approx \mathbf{k}_{2}\approx -\mathbf{k}_{3}\approx -\mathbf{k%
}_{4}.$ These four fermions can then be re-arranged either into a
convolution of two particle-hole bubbles $\Pi _{ph}(\Omega ,q)$ \textit{or}
two particle-particle bubbles~\cite{cm,finn} $\Pi _{pp}(\Omega ,q)$; the
term $\Omega ^{2}/q^{2}$ is produced regardless of the choice.
\begin{figure}[tbp]
\begin{center}
\epsfxsize=1.0\columnwidth\epsffile{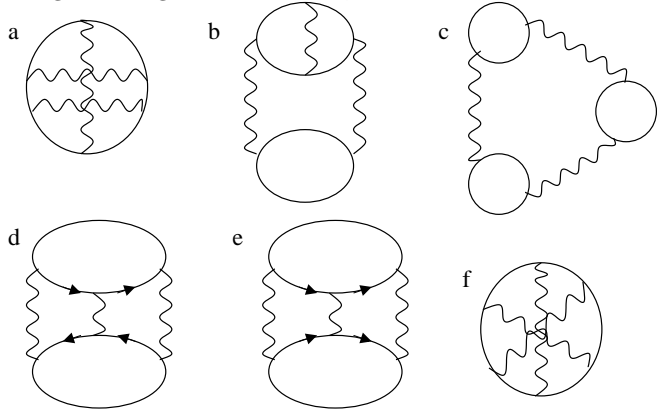}
\end{center}
\caption{Diagrams for the thermodynamic potential to third order in the
interaction.}
\label{fig:omega}
\end{figure}

Relevant third-order diagrams for $\Xi $ are shown in
Fig.\ref{fig:omega}. They contain either three particle-hole or
three particle-particle bubbles. It can be shown that the
non-analytic terms in $\delta \gamma (T)$ come from the terms
containing the products of two dynamic and one static parts of
these bubbles. Two dynamic parts of the bubbles produce $\Omega
^{2}/q^{2}$ term, which is the source of non-analyticity, whereas
the third bubble renormalizes \textit{static} backscattering
vertex. If the third bubble is a particle-hole one, the
renormalized vertex is a constant. The momentum integration then
yields $\Omega ^{2}\ln |\Omega |,$ and subsequent frequency
summation gives rise to the $T^{3}$-term in $\Xi $. In the third
bubble is a particle-particle one, the renormalized vertex
contains an additional factor of $\ln q$. This changes the result
of the momentum integration to $\Omega ^{2}\ln ^{2}|\Omega |$
[note an extra $\ln \left| \Omega \right| $ as compared to
Eq.(\ref{dm1})] and gives rise to a $T\ln T$-term in $\delta
\gamma \left( T\right) .$

The contributions from diagrams (a-d) have already been presented in ~\cite
{cmgg}, diagrams (e) and (f) were not considered there. For completeness, we
present the results for all third-order diagrams. We have~\cite{comm_2}
\begin{widetext}
\begin{eqnarray}
&&\Xi_{3a} = -\left (u_0 \langle u_\theta u_{\pi -\theta}\rangle +
2 u_0 u_{\pi} \langle u_{\theta}\rangle +  u_{\pi}
\langle \langle u^2_{\theta}\rangle\rangle + 2 u_0 u_{\pi} \langle
\langle u_{\theta}\rangle\rangle\right) K,~~
\Xi_{3b} = \left (4 u^2_0 \langle u_\theta\rangle + 2 u^2_0
u_{\pi} \right) K \nonumber \\
&& +\left[ 4 u^2_\pi \langle \langle u_\theta\rangle\rangle +  2
u^2_{\pi} u_0 \langle \langle 1\rangle\rangle \right] K,~~
\Xi_{3c} = - 4 \left[u^3_0  + u^3_{\pi} \langle \langle
1\rangle\rangle\right] K,~\Xi_{3d} = 2 u_{\pi} \langle u_\theta
u_{\pi -\theta}\rangle K +
2 u_0 \langle \langle u^2_\theta\rangle\rangle K, \nonumber \\
&&\Xi_{3e} = -(2 u_0 \langle u^2_\theta\rangle + 2 u_\pi \langle
u_\theta u_{\pi-\theta}\rangle) K \ln{\frac{E_F}{T}}, ~\Xi_{3f} =
( u_0 \langle u_\theta u_{\pi-\theta}\rangle +  u_\pi \langle
u^2_\theta\rangle) K \ln{\frac{E_F}{T}}\label{3e}
\end{eqnarray}
\end{widetext}
where $K\equiv \zeta (3)T^{3}/\pi v_{F}^{2}$ and $u_{\theta }=(m/2\pi
)U(2k_{F}\sin {\theta /2})$, such that $u_{0}=(m/2\pi )U(0)$ and $u_{\pi
}=(m/2\pi )U(2k_{F})$. 
Quantities denoted as $\langle g_{\theta }\rangle $ and $\langle \langle
g_{\theta }\rangle \rangle $ are ($g_{\theta }=u_{\theta },u_{\theta }^{2}$,
and $1$):
\begin{widetext}
\begin{equation}
\langle g_{\theta }\rangle = - \frac{2\pi}{m} \int \frac{d^2 l d \omega}{(2\pi)^3} g(|{\bf l}|) G^2_{{\bf l} + {\bf k}_F} =
\int_{0}^{\pi }\frac{d\theta }{\pi }g_{\theta};~~\langle \langle g_{\theta }\rangle \rangle =  - \frac{2\pi}{m} \int \frac{d^2 l d \omega}{(2\pi)^3} g(|{\bf l}|) G_{{\bf l} + {\bf k}_F} G_{{\bf l} - {\bf k}_F} = \int_{0}^{\pi }\frac{d\theta }{%
\pi }~g_{\theta }~\cos {\frac{\theta }{2}}~\ln \left( \cot \frac{\theta }{4}%
\right) .  \label{10_9_5}
\end{equation}
\end{widetext}
For a circular Fermi surface, $\langle \langle 1\rangle \rangle =1$.

Combining $\Xi_3$ with the second-order result $\Xi _{2}=(u_{0}^{2}+u_{\pi
}^{2}-u_{0}u_{\pi })K$ \cite{cmgg}
we obtain a complete result for $\delta \gamma (T)$ to third order in the
interaction
\begin{widetext}
\begin{eqnarray}
&&\delta\gamma(T) = -\Big[u_{0}^{2}+u_{\pi }^{2}-u_{0}u_{\pi }-
4u^3_0 + 2 u^2_0 u_{\pi} + (4 u^2_0 - 2 u_0 u_{\pi})
  \langle u_\theta\rangle - (u_0 -2u_{\pi})  \langle u_\theta u_{\pi -\theta}\rangle
    + (-4 u^3_\pi + 2 u_0 u^2_\pi) \langle \langle
1\rangle\rangle\notag\\
&& + (4 u^2_\pi - 2 u_0 u_\pi) \langle \langle
u_\theta\rangle\rangle -(u_\pi - 2 u_0) \langle \langle
u^2_\theta\rangle\rangle   -\left(2 u_0 \langle u^2_\theta\rangle
+ 2 u_\pi  \langle u_\theta u_{\pi-\theta}\rangle -
 u_0 \langle u_\theta u_{\pi-\theta}\rangle -  u_\pi \langle u^2_\theta\rangle\right)~
 \ln\frac{E_F}{T}\Big]\frac{6\zeta(3)T}{\pi v_F^2}
\label{k_16}
\end{eqnarray}
\end{widetext}

Next, we compute independently the static interaction vertices $\Gamma (%
\mathbf{k,}\mathbf{k})$ and $\Gamma (\mathbf{k},-\mathbf{k})$ ($\Gamma ^{k}$
in the Fermi liquid notations) to second order in $U(q),$ including
the renormalizations in the particle-hole and particle-particle channels.
Collecting both contributions, we find
\begin{mathletters}
\begin{eqnarray}
\frac{m}{2\pi }\Gamma (\mathbf{k},\mathbf{k}) &=&\tilde{u}_{0}+\langle
\langle u_{\theta }^{2}\rangle \rangle -\langle u_{\theta }^{2}\rangle \ln {%
\frac{E_{F}}{T}};  \label{10_9_6a} \\
\frac{m}{2\pi }\Gamma (\mathbf{k},-\mathbf{k}) &=&\tilde{u}_{\pi }+\langle
u_{\theta }u_{\pi -\theta }\rangle \left( 1-\ln {\frac{E_{F}}{T}}\right) .
\label{10_9_6}
\end{eqnarray}
where
\end{mathletters}
\begin{eqnarray*}
\tilde{u}_{0} &=&u_{0}~\left( 1-2u_{0}+2\langle u_{\theta }\rangle \right) ,
\\
~\tilde{u}_{\pi } &=&u_{\pi }~\left( 1-2u_{\pi }\langle \langle 1\rangle
\rangle +2\langle \langle u_{\theta }\rangle \rangle \right) .
\end{eqnarray*}
The charge and spin components $f_{c,s}(\pi )$ are obtained using Eq. (\ref
{c_1}). Evaluating the expression $f_{c}^{2}(\pi )+3f_{s}^{2}(\pi )$ to
third order in $U(q)$ and substituting the result into Eq.(\ref{jul3_4}), we
find that Eq. ( \ref{k_16}) \textit{is fully reproduced}. Therefore, at
least to third order, the prefactor of the $T$ term in $\delta \gamma (T)$
is expressed via \textit{exact} scattering amplitudes, which include the
renormalizations in the Cooper channel.

At low $T,$ the particle-particle renormalizations are more relevant that
those in the particle-hole channel, as the former contain a large factor of $%
\ln {E_{F}/T}$. Keeping only the particle-particle renormalization, we
obtain
\begin{eqnarray}
f_{c}(\pi ) &=&2u_{0}-u_{\pi }+\left( \langle u_{\theta }u_{\pi -\theta
}\rangle -2\langle u_{\theta }^{2}\rangle \right) \ln {\frac{E_{F}}{T}}
\notag \\
f_{s}(\pi ) &=&-\left( u_{\pi }-\langle u_{\theta }u_{\pi -\theta }\rangle
\ln {\frac{E_{F}}{T}}\right)   \label{4}
\end{eqnarray}
We see that both $f_{s}(\pi )$ and $f_{c}(\pi )$ contain logarithmic
corrections. For $u_{\theta }=$const$\equiv u$, Eqs. (\ref{4}) reduce to
\begin{equation}
f_{c}(\pi )=-f_{s}(\pi )=u\left( 1-u\ln {\frac{E_{F}}{T}}\right) \approx
\frac{u}{1+u\ln {\frac{E_{F}}{T}}}  \label{5}
\end{equation}
A better estimate is obtained if we assume, following AE, a simple model
form for the irreducible interaction in the particle-particle channel: $J$%
\begin{equation*}
J^{C}(q)=\Gamma (k,-k;k+q,-k-q)=aw/(q^{2}+a^{2}),
\end{equation*}
where $q=2k_{F}\sin \theta /2$ and $w$ has the units of velocity. The
partial components of $J^{c}(q)$ are
\begin{equation}
J_{n}^{C}=\frac{we^{-\beta n}}{\sqrt{a^{2}+4k_{F}^{2}}},~~\beta =2\ln \left(
\frac{a}{2k_{F}}{+\sqrt{1+\frac{a^{2}}{4k_{F}^{2}}}}\right)   \label{c_4}
\end{equation}
In the limit of $T\rightarrow 0$, the sums over $n$ in Eq. (\ref{c_2}) for $%
\Gamma ^{C}(0)$ and $\Gamma ^{C}(2k_{F})$ are dominated by large $n$;
replacing summation over $n$ by integration, we obtain
\begin{equation}
\Gamma ^{C}(0)=\frac{2\pi }{m}~\frac{\ln L}{\beta L},~\Gamma ^{C}(2k_{F})=%
\frac{2\pi }{m}~\frac{1-e^{-\beta }}{\beta L},  \label{c_5}
\end{equation}
where $L=\ln {E_{F}/T}$. In this limit, $\Gamma ^{C}(0)$ is larger than $%
\Gamma ^{C}(2k_{F})$ by $\ln L$, hence $f_{c}(\pi )\approx 2\ln L/(\beta
L)\gg f_{s}\left( \pi \right) $, and the full result for the specific heat
coefficient becomes
\begin{equation}
\delta \gamma (T)=-\frac{3\zeta (3)}{2\pi \left( v_{F}^{\ast }\right) ^{2}}~%
\left[ \frac{2\ln {\ln {\frac{E_{F}}{T}}}}{\beta \ln {\frac{E_{F}}{T}}}%
\right] ^{2}T.  \label{c_6}
\end{equation}
Note that the prefactor depends on the functional form of $\Gamma ^{C}(q)$
but not on the magnitude of the interaction. Eq. (\ref{c_5}) is valid only
at such low temperatures that $\ln L\gg 1$. For a more realistic case of $%
L\gtrsim 1$, both $\Gamma ^{C}(0)$ and $\Gamma ^{C}(2k_{F})$ are of order $%
1/L$, and $\delta \gamma (T)\propto T/\left( \ln E_{F}/T\right) ^{2}$.

A more fundamental reason why the ultra-low $T$ regime is
unaccessible is the Kohn-Luttinger effect: the superconducting
instability for a nominally repulsive interaction \cite{kl}. It is
known that irreducible $J^{C}(q)$ is non-analytic near $q=2k_{F}$
due to screening of the original pairing interaction by the
particle-hole excitations. Screening produces a long-range
component of $J^{C}(q)$. In 2D, the Kohn-Luttinger effect is a bit
tricky as one needs to include vertex corrections to the
polarization bubble to obtain an oscillating long-range component
of the pairing interaction $\sin (2k_{F}r{)/}r^{2}$ between
particles at the Fermi surface~ \cite{chub_kl}. Because of
oscillations and $1/r^{2}$ behavior, the partial harmonics
$J_{n}^{C}$ acquire \textit{negative} parts that fall off with $n$
algebraically rather then exponentially: $J_{n}^{C,KL}\approx -\alpha /n^{2}$%
, $\alpha >0$ (at small $U$, $\alpha \propto U^{3}$). As a result, $J_{n}^{C}
$ become negative for $n>n_{c}$, which implies superconductivity. For a
moderately strong interaction [$mU(q)\gtrsim 1$], $J_{n}^{C}$ becomes
negative already for $n=1$ ($|J_{1}^{C}|$ is the largest), and the system
becomes unstable towards $p-$wave pairing at $T_{p}$ defined by $(m\alpha
/2\pi )\ln {E}_{F}/T_{p}=1$. Near $T_{p}$, the $n=1$ term dominates the sums
in Eq. (\ref{c_2}), both $\Gamma ^{C}(0)$ and $\Gamma ^{C}(2k_{F})$ diverge
as $\alpha /(1-(m\alpha /2\pi )L)\propto 1/(T-T_{p})$, hence
\begin{equation}
\delta \gamma (T)=-\frac{36\zeta (3)}{2\pi \left( v_{F}^{\ast }\right) ^{2}}%
~\left( \frac{T_{p}}{T-T_{p}}\right) ^{2}T.  \label{c_7}
\end{equation}

To verify our computational procedure, we also consider the 1D case. The
procedure that we used in 2D is also applicable to the 1D case, with the
only modification that angular averages of the interaction in the
particle-hole channel [Eq.(\ref{10_9_5})] are replaced by a sum of just two
terms, for $\theta =0$ and $\theta =\pi $, as the Fermi surface in 1D
consists of just two points $k=\pm k_{F}$. The integrand in (\ref{10_9_5})
vanishes at $\theta =\pi $ and diverges logarithmically at $\theta =0$,
i.e., for $D=1$, the renormalization of the interaction in the particle-hole
channel also leads to logarithmic corrections:
\begin{equation}
\langle \langle u_{\theta }\rangle \rangle \rightarrow u_{0}\ln {\frac{E_{F}%
}{T}},~~\langle \langle u_{\theta }^{2}\rangle \rangle \rightarrow
u_{0}^{2}\ln {\frac{E_{F}}{T}},~~\langle \langle 1\rangle \rangle
\rightarrow \ln {\frac{E_{F}}{T}},~~  \label{new_1}
\end{equation}
where now $u_{\theta }=U(2k_{F}\sin \theta /2)/(2\pi v_{F})$.
Combining the logarithms in the particle-particle and
particle-hole channel and neglecting non-logarithmic second-order
terms, we reproduce the well-known results for the charge- and
spin scattering amplitudes in 1D \cite{bychkov,review,emery}:

\begin{equation*}
f_{c}(\pi )=2u_{0}-u_{\pi },~f_{s}(\pi )\approx -\frac{u_{\pi }}{1+2u_{\pi
}\ln {\frac{E_{F}}{T}}}
\end{equation*}
The logarithmic corrections are cancelled out in $f_{c}(\pi )$, but are
present in $f_{s}(\pi )$.  The interplay between the behavior of the
renormalized spin amplitude and the specific heat in 1D is a more subtle
issue, because the backscattering part of $\delta \gamma (T)$ in 1D contains extra $O(T)$ terms and
does not reduce to Eq. (\ref{jul3_4}) with the renormalized $f_{s}(\pi )$~
\cite{ae,lukyanov,comm_new}.

To elucidate the difference between  2D and 1D further, we
consider a
2D system with a non-circular Fermi surface. Near an arbitrary point $%
\mathbf{k}_{F}$ on such a surface, the fermionic dispersion can be expanded
as
\begin{equation}
\epsilon _{\mathbf{k}}=v_{F}k_{||}+\frac{k_{\perp }^{2}}{2m_{c}},
\label{curv}
\end{equation}
where $k_{||}$ and $k_{\perp }$ are the projections of vector $\mathbf{k-k}%
_{F}$ on the normal and tangent to the Fermis surface at point
$\mathbf{k}_{F}
$, correspondingly, $v_{F}$ is the local value of the Fermi velocity, and $%
m_{c}=1/\kappa v_{F}$ is related to the local curvature, $\kappa ,$ of the
Fermi surface \cite{chm,comm_curv}.   For a circular Fermi surface, $\kappa
=k_{F}^{-1}$ so that $m_{c}=m=k_{F}/v_{F}$ and our 2D results are valid. If $%
m_{c}\gg m$, the dispersion near two symmetric points $\pm
\mathbf{k}_{F}$ is almost one-dimensional and, if only $u_{0}$ and
$u_{\pi }$ are relevant, we should reproduce 1D results. Indeed,
evaluating  the particle-hole contributions, labelled above as
$\langle \langle ...\rangle \rangle $, for the case of $m\gg
m_{c},$ we find that they have the same logarithmic behavior, as
in Eq.(\ref{new_1}); the only difference being that the
logarithm is now cut by the largest of the two energies: $T$ and $%
E_{c}=k_{F}^{2}/2m_{c}$. Substituting these results into the expression for $%
f_{c}(\pi )$, we find that to logarithmic accuracy
\begin{equation}
f_{c}(\pi )=2u_{0}-u_{\pi }-2(u_{0}^{2}+u_{\pi }^{2}-u_{0}u_{\pi })\ln {max(E%
}_{c}{/T,1)}
\end{equation}
For $m_{c}=\infty $, $E_{c}=0$, and the logarithmic term vanishes,

 just as it happens in 1D.

We see that the curvature of the Fermi surface is the crucial element of 
2D consideration~\cite{khvesh}. When the curvature is finite, the particle-hole renormalizations of the scattering ampludes and of $\delta \gamma (T)$ are not logatithmic at the smallest $T$, and the logarithms only come from particle-particle renormalizations. Then  $f_s (\pi)$, $f_s (\pi)$, and $\delta \gamma (T)$  are all logarithmically reduced.  When the curvature is zero, the dispersion is one-dimensional, particle-hole renormalizations
 also become logarithmic, and for $f_c (\pi)$ the logarithms from the 
particle-particle and particle-hole renormalizations are cancelled out. 
In the eikonal approximation used by AE and in earlier 2D bosonization theories~\cite{fradkin}, the curvature of the Fermi surface is neglected. In this situation, the charge amplitude behaves as in 1D and is not renormalized, and $\delta \gamma (T)$  remains linear in $T$.

To summarize, in this paper we analyzed the effect of the Cooper-channel
renormalization on the temperature dependence of the specific heat. We have
shown the non-analytic term in the specific heat coefficient of a 2D Fermi
liquid is expressed via the square of the \textit{full }backscattering
amplitude $f(\pi )$, renormalized in both particle-hole and
particle-particle channels. Due to the particle-particle renormalization,
both the charge and spin components of $f(\pi )$ are reduced by a factor of $%
\ln (E_{F}/T)$ in the limit of $T\rightarrow 0.$  Consequently,
the temperature dependence of $\delta \gamma (T)$ is $TS\left(
T\right) /\ln ^{2}(E_{F}/T)$, where $S\left( T\right) $ is a
slowly varying function, whose form depends on the details of the
interaction in the particle-particle channel. When applied to 1D,
our method reproduces the cancellation between particle-hole and
particle-particle contributions to the charge channel. The
logarithmic renormalization of the charge amplitude is due to a
specifically higher-dimensional effect--a finite curvature of the
Fermi surface. For a flat Fermi surface, this renormalization is
absent.

We acknowledge helpful discussions with I. L. Aleiner, A. M. Finkelstein, L.
I. Glazman, K. B. Efetov, A. A. Nersesyan, R. Saha and G. Schwiete, support
from NSF-DMR 0604406 (A. V. Ch.), 0308377 (D. L. M.), and the hospitality of
the Aspen Center of Physics.


\begin{thebibliography}{99}
\bibitem{FL}  C.J. Pethick and G.M. Carneiro, Phys. Rev. A \textbf{7}, 304
(1973) and references therein.

\bibitem{belitz}  D. Belitz, T. R. Kirkpatrick, and T. Vojta, Rev. Mod.
Phys. \textbf{77}, 579 (2005) and references therein.

\bibitem{all_1}  D. Coffey and K. S. Bedell, Phys. Rev. Lett. \textbf{71},
1043 (1993).

\bibitem{all_2}
 M. A. Baranov, M. Yu. Kagan, and M.
S. Mar'enko, JETP Lett. \textbf{58}, 709 (1993).

\bibitem{all_3} G. Y. Chitov and A. J. Millis, \prl {\bf 86}, 5337 (2001).


\bibitem{cm}  A. V. Chubukov and D. L. Maslov, Phys. Rev. B \textbf{68, }%
155113 (2003); Phys. Rev. B \textbf{74}, 079907 (2006).

 \bibitem{all_4} J.
Betouras, D. Efremov, and A. Chubukov, Phys. Rev. B \textbf{72},
115112 (2005).

\bibitem{efetov}  G. Schwiete and K. B. Efetov, Phys. Rev. B \textbf{74}, 165108
(2006).

\bibitem{finn}  A. Shekhter and A.M. Finkelstein, Phys. Rev. B \textbf{74}, 205122
(2006); Proc. Nat. Acad. Sci. \textbf{103} (2006) 15765; Proc.
Nat. Acad. Sci. \textbf{103} (2006) 18874.

\bibitem{we}  D. L. Maslov, A. V. Chubukov, and R. Saha, Phys. Rev. B \textbf{74}, 220402 (2006).

\bibitem{cmgg}  A. V. Chubukov, D. L. Maslov, S. Gangadharaiah, and L. I.
Glazman, Phys. Rev. Lett. \textbf{95}, 026402 (2005); Phys. Rev. B
\textbf{71}, 205112 (2005).

\bibitem{cmm}  A. V. Chubukov, D.L. Maslov, and A. J. Millis, Phys. Rev. B
\textbf{73}, 045128 (2006).

\bibitem{chm} A.V. Chubukov and A.J. Millis, Phys. Rev. B \textbf{74}, 115119 (2006).

\bibitem{comm_a}  The spin susceptibility both in 2D and 3D, as well as the
specific heat in 3D, contain other contributions which do not reduce to 1D
scattering~\cite{finn,we}].

\bibitem{ae}  I. L. Aleiner and K. B. Efetov, Phys. Rev. B \textbf{74},
075102 (2006); cond-mat/0610345. Notice that the ``backscattering amplitude''
is defined in these papers as the irreducible amplitude, without the renormalizations in the Cooper
channel.

\bibitem{bychkov}  Yu. A. Bychkov, L. P. Gor'kov, and I. E. Dzyaloshisnkii,
Sov.\ Phys.\ JETP \textbf{23}, 489 (1966).

\bibitem{review}  J. Sol{\'o}yom, Adv. Phys. \textbf{28}, 209 (1979).

\bibitem{emery}  V. J. Emery, in \textit{Highly Conducting One-Dimensional
Solids}, eds. J. T. Devreese, R. E. Evrard, and V. E. van Doren, (Plenum
Press, New York, 1979), p. 247.

\bibitem{schulz_95}  H. J. Schulz, in \textit{Mesoscopic Quantum Physics},
Les Houches XXI (eds. E. Akkermans, G. Montambaux, J. L. Pichard, and
J. Zinn-Justin)
, (Elsevier, Amsterdam, 1995), p. 533; 
G. I. Japaridze and A. A. Nersesyan, Phys. Lett.\textbf{\ 94} A, 224 (1983).

\bibitem{landau}  E.M. Lifshitz and L.P. Pitaevskii, Statistical Physics,
p.2, Pergamon Press, 1996.

\bibitem{comm_2}  We corrected a sign error in $\Xi _{3a}$ in Ref. \cite
{cmgg}.

\bibitem{kl}  W. Kohn and J. Luttinger, Phys. Rev. Lett. 15, (1965).

\bibitem{chub_kl}  A. Chubukov, Phys. Rev. B 48, 1097-1104 (1993)

\bibitem{lukyanov}  S. Lukyanov, Nucl. Phys. B \textbf{522, }533 (1998).

\bibitem{comm_new}  We thank I. Aleiner and K. Efetov for pointing out a
mistake in our original treatment of the 1D case.


\bibitem{comm_curv}  Out of all second-order terms  we
keep in Eq.(\ref{curv}) the one that can be comparable to the first-order term ($%
v_{F}k_{||}).$

\bibitem{khvesh} For 
a related discussion on the role 
of the Fermi surface curvature in 2D, see A. Chubukov 
and D. Khveshchenko, Phys. Rev. Lett.  {\bf 97}, 226403  (2006).

\bibitem{fradkin} see. e.g.,  M. J. Lawler et al.,
 Phys. Rev. {\bf B73}, 085101 (2006) and references therein.

\end{thebibliography}
\end{document}